\RequirePackage{fix-cm}
\documentclass[pdftex,twocolumn,epjc3,final]{svjour3}  
\smartqed  % flush right qed marks, e.g. at end of proof
\RequirePackage[T1]{fontenc}
\RequirePackage{xcolor}
\RequirePackage{amssymb}
\RequirePackage{amsmath}
\RequirePackage{graphicx} 
\RequirePackage{epstopdf} 
\newcommand*{\Apr}{{A^\prime}}
\RequirePackage[colorlinks,citecolor=black,urlcolor=black,linkcolor=black]{hyperref}
\journalname{Eur. Phys. J. A}
\begin{document}

\title{Light dark matter searches with positrons}

\titlerunning{LDM with $e^+$@JLab}   % if too long for running head

%\thankstext{Cp}{Contact person: }

\author{
M.~Battaglieri\thanksref{addr1,addr2}
\and
A.~Bianconi\thanksref{addr3,addr4}
\and
P.~Bisio\thanksref{addr5}
\and
M.~Bond\`i\thanksref{addr1}
\and
A.~Celentano\thanksref{addr1}
\and 
G.~Costantini\thanksref{addr3,addr4}
\and
P.L.~Cole\thanksref{addr16}
\and
L.~Darm\'e\thanksref{addr11}
\and
R.~De~Vita\thanksref{addr1}
\and
A.~D'Angelo\thanksref{addr7,addr8}
\and
M.~De~Napoli\thanksref{addr9}
\and
L.~El~Fassi\thanksref{addrMiss}
\and
V.~Kozhuharov\thanksref{addr11,addr10}
\and
A.~Italiano\thanksref{addr9}
\and
G.~Krnjaic\thanksref{addr14,addr15}
\and
L.~Lanza\thanksref{addr7}
\and
M.~Leali\thanksref{addr3,addr4}
\and
L.~Marsicano\thanksref{addr1,Cp}
\and
V.~Mascagna\thanksref{addr4,addrUniins}
\and
S.~Migliorati\thanksref{addr3,addr4}
\and
E.~Nardi\thanksref{addr11}
\and
M.~Raggi\thanksref{addr12,addr13,Cp}
\and
N.~Randazzo\thanksref{addr9}
\and
E.~Santopinto\thanksref{addr1}
\and
E.~Smith\thanksref{addr2}
\and
M.~Spreafico\thanksref{addr5}
\and
S.~Stepanyan\thanksref{addr2}
\and
M.~Ungaro\thanksref{addr2}
\and
P.~Valente\thanksref{addr13}
\and
L.~Venturelli\thanksref{addr3,addr4}
\and
M.H.~Wood\thanksref{addrCanis}
}

\institute{Istituto Nazionale di Fisica Nucleare, Sezione di Genova, 16146 Genova, Italy
\label{addr1}
\and
Thomas Jefferson National Accelerator Facility, Newport News, Virginia 23606
\label{addr2}
\and
Universit\`a  degli  Studi  di  Brescia,  25123  Brescia,  Italy
\label{addr3}
\and
INFN,  Sezione  di  Pavia,  27100  Pavia,  Italy
\label{addr4}
\and
Universit\`a degli Studi di Genova, 16146 Genova, Italy
\label{addr5}
\and
Lamar  University,  4400  MLK  Blvd,  PO  Box  10046,  Beaumont,  Texas  77710
\label{addr16}
\and
Istituto Nazionale di Fisica Nucleare, Laboratori Nazionali di Frascati, Via E. Fermi 54, Frascati, Italy
\label{addr11}
\and
INFN,  Sezione  di  Roma  Tor  Vergata,  00133  Rome,  Italy
\label{addr7}
\and
Universit\`a  di  Roma  Tor  Vergata,  00133  Rome  Italy
\label{addr8}
\and
Istituto Nazionale di Fisica Nucleare, Sezione di Catania, 95125 Catania, Italy
\label{addr9}
\and
Mississippi State University, Mississippi State, Mississippi 39762-5167, USA
\label{addrMiss}
\and
Faculty of physics, University of Sofia, 5 J. Bourchier Blvd., 1164 Sofia, Bulgaria
\label{addr10}
\and
Fermi National Accelerator Laboratory, Batavia, Illinois 60510, USA
\label{addr14}
\and
Kavli Institute for Cosmological Physics, University of Chicago, Chicago, Illinois 60637, USA
\label{addr15}
\and
Università degli Studi dell'Insubria, 22100 Como, Italy
\label{addrUniins}
\and
Sapienza Universit\`a di Roma, piazzale Aldo Moro 5 Roma, Italy
\label{addr12}
\and
Istituto Nazionale di Fisica Nucleare, Sezione di Roma, piazzale Aldo Moro 5 Roma, Italy
\label{addr13}
\and
Canisius College, Buffalo, NY 14208, USA
\label{addrCanis}
}

\thankstext{Cp}{Contact authors: luca.marsicano@ge.infn.it, \\ mauro.raggi@roma1.infn.it}

\authorrunning{M.~Battaglieri {\it et al.}} % if too long for running head

\date{Draft : \today}

\maketitle

\begin{abstract}

We discuss two complementary strategies to search for light dark matter (LDM)  exploiting the 
posi\-tron beam possibly available in the future at Jefferson Laboratory.
% this facility. 
LDM is a new compelling hypothesis that identifies dark matter with new sub-GeV ``hidden sector'' states, neutral under standard model interactions and interacting with our world through a new force. Acceler\-ator-based searches at the intensity frontier are uniquely suited to explore it. Thanks to  
the high intensity and the high energy 
% to the unique properties 
of the CEBAF (Continuous Electron Beam Accelerator Facility) beam,  
% -- the high intensity and the high energy -- 
and relying on a novel LDM production mechanism via positron annihilation on target 
atomic electrons, the proposed strategies will allow us to explore new regions in the LDM 
parameters space, 
% confirming or ruling out 
thoroughly probing the LDM hypothesis as well as more general hidden sector scenarios. 
%
% \keywords{...}
%
\end{abstract}

%\hyphenation{mo-del}

\vspace{-5mm}

\section{Introduction and motivations}

% \subsection{Overview}
One of the most compelling % empirical 
arguments motivating the search for extensions of the Standard Model (SM) 
 is the need to explain the nature of dark matter (DM). In years past, theoretical 
and experimental efforts mainly catalysed around the hypothesis that DM corresponds to a Weakly 
Interacting Massive Particle (WIMP) with electroweak scale mass.
%   (see e.g. Ref. [1] for a recent review). 
Such a hypothesis is  well ground\-ed since in the early Universe
WIMPs would be produced via thermal processes, and their  annihilation 
with typical weak interaction rates would leave, almost independently of other details, 
a relic density of the correct size to explain DM observations. 
However,  despite an extensive % and long lasting 
search program that combined 
direct, indirect,  and collider probes,  to date no conclusive signal of the WIMP has been 
found~\cite{Arcadi:2017kky}. This prompts the scientific community  to put no lesser vigor 
in exploring  alternative pathways. 

Feebly interacting particles (FIPs) represent a particularly interesting alternative to WIMPs.  
In recent years the physics of FIPs has focused a steadily growing interest,  as characterised witnessed by the remarkable 
 number of  community reports and white papers that have appeared in the last few 
 years~\cite{Feng:2014uja,Hewett:2014qja,Alexander:2016aln,Battaglieri:2017aum,Beacham:2019nyx,Agrawal:2021dbo}.
FIPs are  exotic and relatively light  particles, not charged under the SM % $SU(3)_C \times SU(2)_L \times U(1)_Y$ 
gauge group, whose interactions with the SM fields are  extremely suppressed. FIPs are often assumed  
to be part of a possibly complicated  secluded sector,  called the {\it dark sector}, with    
the Lightest stable dark particle(s) playing the role of DM (LDM).   
This scenario has sound theoretical motivations: in first place many known particles are 
uncharged under some  gauge group factors, so  that the existence of particles blind 
to all  $SU(3)_C \times SU(2)_L \times U(1)_Y$ interactions seems a rather natural possibility. 
Secondly,  theoretical mechanisms like gauge symmetries or quasi-exact spontaneously broken  global 
symmetries, that we know are realised in Nature,  can explain why some particles remain light even 
when they are associated with  physics at some large scale.
% a zero-mass photon is generated upon spontaneous breaking of the electroweak symmetry associated  
% with   \mbox{$\Lambda_{\rm EW} \sim 100\,$GeV},   pions are the Goldstone-bosons of a global 
% symmetry broken spontaneously at a scale $\Lambda_{\rm QCD} \gg m_\pi$.  
% %
From the phenomenological point of view, light weakly coupled new particles have been often 
invoked to account for several discrepancies % with SM predictions 
 observed in low energy experiments. 
% between low energy experimental results and SM predictions. 
Examples are the measured value of the muon
anomalous magnetic moment~\cite{Blum:2013xva}, the proton
charge radius measured in muonic atoms~\cite{Pohl:2010zza,Carlson:2015jba,Krauth:2017ijq}, 
the discrepancy between neutron lifetime measurements
in bottle and beam experiments~\cite{Wietfeldt:2011suo,Greene:2016aa},
the  measured abundance of primordial $^7$Li  which is 
  a factor of three lower than  BBN predictions~\cite{Sbordone_2010}, 
 the   `bump'  in the  angular distribution of $e^{\pm}$  pairs  observed 
  by  the Atomki collaboration 
 in  nuclear decays  of  $^8$Be~\cite{PhysRevLett.116.042501} and $^4$He~\cite{Krasznahorkay:2019lyl}
excited states.

FIPs scenarios hint to a remarkably broad range of possibilities, ranging from the very nature of 
the new  particles  (scalars, pseudoscalars, fermions, spin-one bosons) and spanning several 
order of magnitude in mass and couplings.  To thoroughly explore all these possibilities will require an  
extensive collaboration among   a variety   of small/medium scale experiments,  exploiting diversified  
detection techniques, and operating at different facilities. 
Accelerator-based searches exploiting positron beams stand out as a particularly powerful tool. This is 
because for any given beam energy there is a range of masses where dark bosons can be produced resonantly
through positron annihilation on atomic electrons in the target, yielding a huge enhancement in the 
production rate.   As it was first pointed out in Ref.~\cite{Nardi:2018cxi}, for resonant production 
% there is no need that the beam energy will be tuned to match in the c.d.m. the precise value of the 
there is no need that the beam energy will be tuned to match in the c.m. the precise value of the 
mediator mass. This is because due to the continue loss of energy from soft photon bremsstrahlung, in the 
first few radiation lengths of a dump a positron beam can continuously scan for production of new resonances. 
While it was later recognised that this process  is of more general importance since, due  
to  the presence  of secondary positrons in EM showers, it contributes to FIPs production also in   electron~\cite{Marsicano:2018krp,Marsicano:2018glj} and proton~\cite{Celentano:2020vtu}  beam 
experiments, it is clear that the availability of a beam of primary positrons is of utmost 
importance to fully take advantage of the resonant production channel.

\subsection{Dark sectors and relic density targets}

Experimental proposals aiming at detecting missing energy from the apparatus are mostly sensitive to the FIPs nature and couplings. While this allows such search strategies to cover a broad class of new light physics models, the details of the dark sector to which the DM belongs remain ultimately inaccessible. Relic density targets thus must rely on additional theoretical assumptions. Arguably the most elegant models for LDM are those which reproduce the successes of WIMP models, being: UV-insensitive, predictive and as economical as possible. A generic expectation is then that DM annihilation proceeds via a bosonic feebly-interacting mediator,  taking care 
that  by the time of CMB annihilation is sufficiently suppressed to evade standard limits on energy 
injection~\cite{Slatyer:2009yq,Slatyer:2015jla,Aghanim:2018eyx}. 

Various simplified structures for the dark sector itself have been considered throughout the years, ranging from basic scenarios such as a complex scalar DM to more advanced setups as asymmetric DM or inelastic DM with an additional dark Higgs boson. The broad target region where $\Omega_{\rm DM}h^2 = 0.11933 \pm 0.00091$~\cite{Aghanim:2018eyx} corresponds to mediator FIPs in the $1 \,  \rm MeV  - 10 \, \, \rm GeV$ mass window  with coupling suppressed as $10^{-5} - 10^{-2}$ respectively. However, the precise relic density target strongly depends on the details of the model. Assuming for simplicity a dark photon $\Apr$ scenario, interaction with the SM  can proceed via a kinetic mixing $\varepsilon$. The dark photon couples with 
the electromagnetic current $\mathcal{J}_{\rm{em}}^{\mu}$ with coupling $e \varepsilon$ and with the dark gauge current $\mathcal{J}_{\rm{D}}^{\mu}$ leading to:
\begin{eqnarray}
     \mathcal{L} \supset - \Apr_\mu e \varepsilon \mathcal{J}_{\rm{em}}^{\mu} - \Apr_\mu g_D \mathcal{J}_{\rm{D}}^{\mu} \ .
\end{eqnarray}
Different DM candidates corresponds to different choices of dark currents (see e.g. a summary in~\cite{Berlin:2018bsc}) and different thermal targets:
\begin{equation}
    \label{eq:darkcurrent}
    \mathcal{J}_{D}^{\mu} = \begin{cases}
    i \left( \chi^* \partial^\mu \chi - \chi \partial^\mu \chi^* \right) \ ( \rm Complex \ Scalar \ DM, \chi) \\[0.5em]
     \frac{1}{2} \bar \chi \gamma^\mu \gamma^5 \chi \ ( \rm Majorana \ DM, \chi)  \\[0.5em]
     i \bar \chi_1 \gamma^\mu \chi_2 \ ( \rm pseudo-Dirac \ DM, \chi_1)\,.
    \end{cases} \nonumber
\end{equation}
In the simplest cases where the DM relic density is fixed completely via the freeze-out of a dark photon-induced $s$-channel annihilation, the final relic density depends only on
one variable that, for $m_\chi \ll m_{\Apr}$, reads:
\begin{equation}
y \equiv \varepsilon^2 \alpha_D \left(\frac{m_\chi}{m_{\Apr}} \right)^4   \ .
\end{equation}
Depending on the DM nature the typical values of $y$ required to match the relic density target 
can vary by a few orders of magnitude.
% as we will illustrate in the result section. 
A simple example of the effect of the dark sector structure on the relic density is found in the inelastic DM scenario (see~\cite{TuckerSmith:2001hy,Izaguirre:2015zva} and subsequent literature). The dominant annihilation channel at low masses, $\chi_1 \chi_2 \to \Apr^* \to e^+ e^-$, depends  on the mass splitting between the states (co-annihilation mechanism~\cite{Griest:1990kh,Izaguirre:2017bqb}) as well as on whether the mediator can be produced resonantly when $m_{\chi_1}+m_{\chi_2} \lesssim m_{\Apr}$ (resonant annihilation mechanism~\cite{Griest:1990kh,Feng:2017drg}). The former suppresses exponentially the annihilation rate, thus greatly increasing the thermal target in $y$, while the latter  enhances the annihilation, decreasing the thermal target. Finally, the presence of a dark Higgs boson, typically required in UV realisation of this scenario can  open additional annihilation channels~\cite{Choi:2016tkj,Darme:2017glc,Darme:2018jmx}.

% \vspace{0.4cm}

Although LDM models represent a particularly interesting  target, the experimental setups described in this work can be used more generally to search for a large range of FIPs. In particular the limits shown in the case of a dark photon straightforwardly apply to any invisibly-decaying vector boson with coupling to electron/positron $g_e$ with the matching $g_e \leftrightarrow e \varepsilon$. In the case of a (pseudo)-scalar mediator (resp. axion-like particle) with coupling $y_e$ (resp. $g_{ae} m_e$) we have, away from the kinematic threshold, the approximate equivalence $e \epsilon \leftrightarrow y_e / \sqrt{2} \leftrightarrow g_{ae} m_e / \sqrt{2}$ where the $\sqrt{2}$ arises from the different number of degrees of freedom.

\section{Dark sector searches with positron beams on fixed targets}

%%%%%%%%%%%%%%%%%%%%%%%%%%%%%
\begin{figure}[t!]
\begin{center}
\includegraphics[scale=0.45]{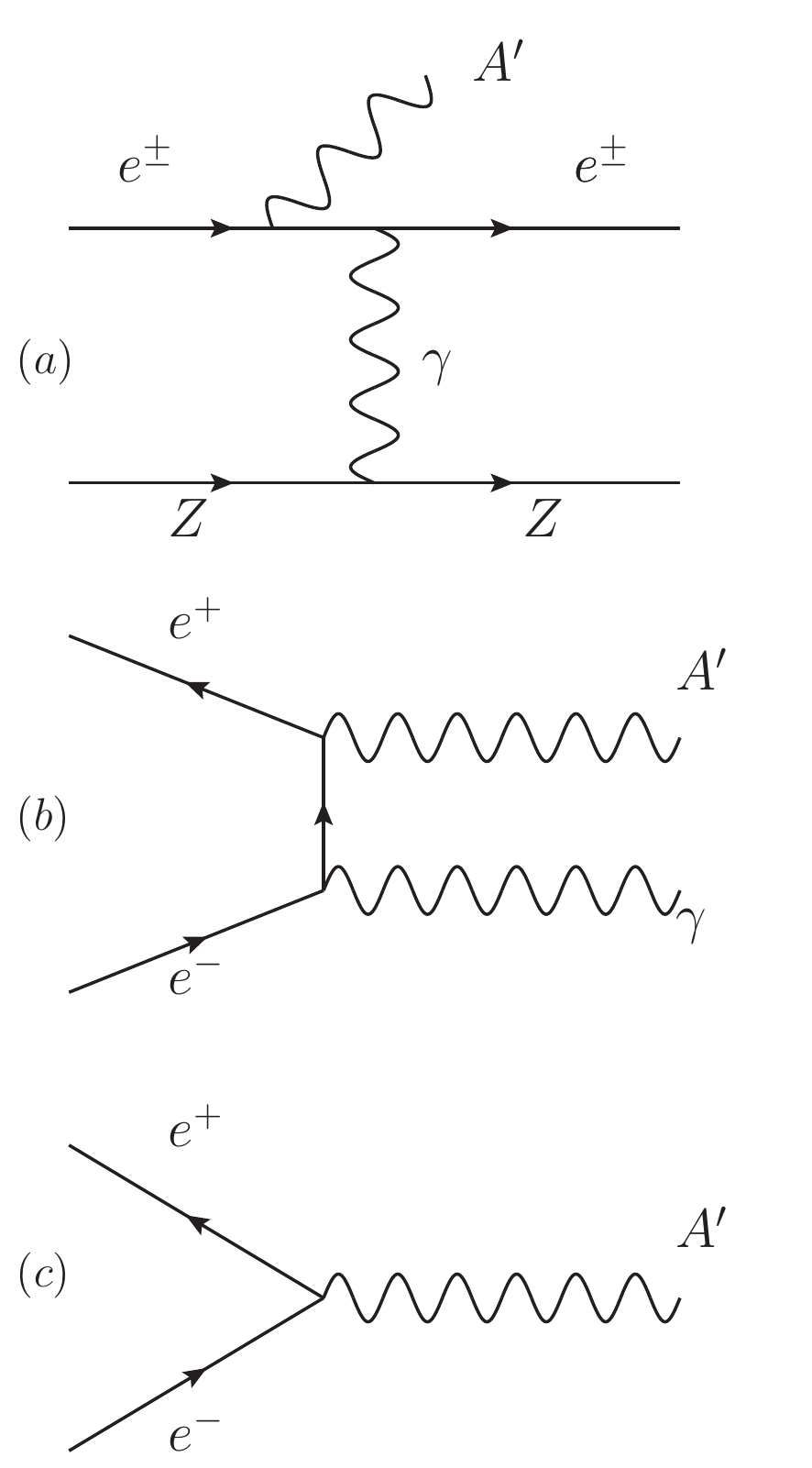}
%\includegraphics[width=0.60\linewidth,height=3cm]{diagram-1}
%\vspace{0.5cm}
%\includegraphics[width=0.60\linewidth,height=3cm]{diagram-2}
%\vspace{0.5cm}
%\includegraphics[width=0.60\linewidth,height=3cm]{diagram-3}
\caption{Three different $\Apr$ production modes in fixed target lepton beam experiments: 
  $(a)$~$\Apr$-strahlung in $e^-/e^+$-nucleon scattering;
  $(b)$~$\Apr$-strahlung in $e^+e^-$ annihilation; 
  $(c)$~resonant $\Apr$ production in $e^+e^-$ annihilation.}
\label{fig-mech}
 \end{center}
 \vspace{-0.5cm}
\end{figure}
%
%%%%%%%%%%%%%%%%%%%%%%%%%%%%%%
The production of LDM particles can be generated in collisions of electrons or positrons of several GeV with a fixed target by the processes depicted in Fig.~\ref{fig-mech}, with the final state $\Apr$ decaying to a $\chi\overline{\chi}$ pair.
For experiments with electron beams, diagram $(a)$, analogous to ordinary photon bremsstrahlung, is the dominant process, although it was recently shown that for thick-target setups, where positrons are generated as secondaries from the developing electromagnetic shower, diagrams $(b)$ and $(c)$ give non-negligible contributions for selected regions of the parameters space~\cite{Marsicano:2018glj} -- see Ref.~\cite{Battaglieri:2017aum} for a comprehensive review of past/current experiments and future proposals. 
On the other hand, for experiments with positron beams, diagrams $(b)$ and $(c)$ play the most important role. %Depending on the target, different approaches are possible.
In this document, we present two complementary measurements to search for light dark matter with positron beams at Jefferson Laboratory, exploiting the unique potential of the proposed $e^+$-beam facility. In the following, we introduce the two approaches and, for each one, we briefly discuss the experimental setup, the measurement strategy, the data analysis and the envisioned results. We underline that Jefferson Laboratory is playing a leading role in the LDM searches, with different experiments already running, HPS~\cite{Celentano:2014wya} and APEX~\cite{Franklin:2017ndz}, or approved to run in the near future, BDX~\cite{Marsicano:2018icx} and DarkLight~\cite{Corliss:2017tms}. %\\ [-30pt]

\subsection{Thin-target measurement}  This measurement exploits the \textit{$\Apr$-strahlung} production in electron-positron annihilation described by diagram $(b)$. The primary positron beam impinges on a thin target, where a photon-$\Apr$  is produced. By detecting the final-state photon in an electromagnetic calorimeter, the missing mass kinematic variable $M_{miss}$ can be computed event-by-event:
\begin{equation}
\label{eq:missmass}
    M^2_{miss}=(P_{beam}+P_{target}-P_\gamma)^2 \; \; .
\end{equation}
The signal would show up as a peak in the missing mass distribution, centered at the $\Apr$ mass, on top of a smooth background due to SM processes resulting from events with a single photon measured in the calorimeter. The peak width is mainly determined by the energy and angular resolution of the calorimeter. Several experiments searching for $\Apr$ with this approach have been proposed.  PADME (Positron Annihilation into Dark Matter Experiment) at LNF~\cite{Raggi:2014zpa} is one of the first $e^+$ on thin target experiment searching for $\Apr$. It uses the 550~MeV positron beam provided by the $DA \Phi NE$ linac at INFN LNF (Laboratori Nazionali di Frascati) impinging on a thin diamond target.

\subsection{Active thick-target measurement} This measurement exploits the \textit{resonant $\Apr$ production} by positrons annihilation on atomic electrons described by diagram $(c)$. The primary positron beam impinges on a thick active target, and the \textit{missing energy} signature of produced and undetected $\chi$ is used to identify the signal~\cite{PhysRevD.91.094026}. The active target measures the energy deposited by the individual beam particles: when an energetic $\Apr$ is produced, its {\it invisible} decay products -- the $\chi\overline{\chi}$ pair -- will carry away a significant fraction of the primary beam energy, thus resulting in measurable reduction in the expected deposited energy. Signal events are identified when the \textit{missing energy} $E_{miss}$, defined  as the  difference between the beam energy and the detected energy, exceeds a minimum threshold value. The signal has a very distinct dependence on the missing energy
%the measured observable, since the corresponding missing energy is kinematically constrained by the dark photon mass,
through the relation\footnote{$m_\Apr$ is the dark photon mass and $m_e=0.511$ MeV/c$^2$ is the electron mass.} $m_\Apr=\sqrt{2m_e E_{miss}}$. This results in a specific experimental signature for the signal, that would appear as a peak in the missing energy distribution, at a value depending solely on the %unknown
dark photon mass. Thanks to the emission of soft bremsstrahlung photons, the thick target provides an almost continuous energy loss for the impinging positrons.
Even though the positron energy loss is a quantized process, the finite intrinsic width of the dark photon -- much larger than the positron energies differences -- and the electrons energy and momentum spread induced by atomic motions~\cite{Nardi:2018cxi} will indeed compensate this effect.  This allows the primary beam to ``scan'' the full range of dark photon masses from the maximum value (corresponding to the loss of all the beam energy), to the minimum value fixed by the missing energy threshold~\cite{Marsicano:2018krp}, exploiting the presence of secondary positrons produced by the developing electromagnetic shower.

\section{Positron annihilation on a thin target}

%\textcolor{red}{LUCA: da JPOS}

\subsection{Signal signature and yield}

The differential cross section for dark photon production via the positron annihilation on the atomic electron of the target $e^+ e^- \rightarrow A' \gamma$, is given by:  
\[  \frac{d \sigma}{d z} = \frac{4 \pi \alpha^2 \varepsilon^2}{s}  \left( \frac{s-m_{A'}^2}{2s} \frac{1+z^2}{1-\beta^2 z^2}  +  \frac{m_{A'}^2}{s-m_{A'}^2}\frac{1}{1-\beta^2 z^2} \right).  \]
Here   $s$ is the $e^+$ $e^-$ system invariant mass squared, $z$ is the cosine of the $A'$ emission angle in the CM frame, measured with respect to the positron beam axis, and $\beta = \sqrt{1- \frac{4 m_e^2}{s}}$. This result has been derived at tree level, keeping the leading $m_e$ dependence to avoid non-physical divergences when $|z| \rightarrow 1$. The emission of the annihilation products in the CM frame is concentrated in the $e^+ e^-$ direction. This results in an angular  distribution for the emitted $\gamma$  peaked in the forward
direction  in the laboratory frame. % the effect being more intense for large values of $m_{\Apr}$. % The emitted $\Apr$ energy distribution ranges from the resonant to the primary positron energy. 
In the case of invisible decays, the $\Apr$ escapes detection, while the $\gamma$ can be detected in the downstream electromagnetic calorimeter (ECAL). The measurement of the photon energy and emission angle, together with the precise knowledge of the primary positron momentum, allows computing the missing mass kinematic variable from Eq.~\ref{eq:missmass}. The mass range that can be spanned is constrained by the available energy in the center of mass frame: using an 11 GeV positron beam at JLab, $\Apr$ masses up to $\sim 106$ MeV$/c^2$ can be explored.
The signal yield  has been evaluated using CALCHEP~\cite{alex2004calchep}; the widths $\sigma(m_{\Apr})$ of the missing mass distributions of the measured recoil photon  has been computed for six different values of the $\Apr$ mass value in the 1--103 MeV range. CALCHEP provides the total cross section of the process, for $\varepsilon = 1$; the cross section value as a function of $\varepsilon$ has been obtained multiplying it by $\varepsilon^2$. 
\begin{figure}[t!]
    \centering
     \includegraphics[width=.5\textwidth]{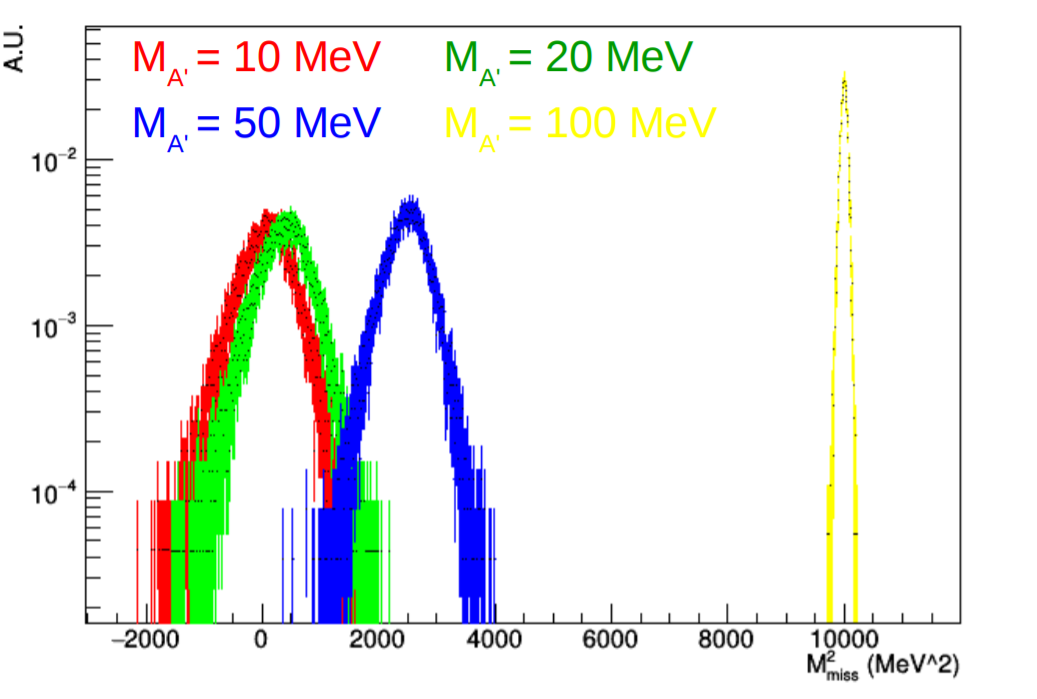}
    \caption{\label{fig:MMres} Computed missing mass spectrum for signal events for 4 different values of $m_\Apr$.}
\end{figure}
Figure~\ref{fig:MMres} shows results for 4 mass values: due to the $e^+ e^- \rightarrow \gamma \Apr$ process kinematics, the missing mass resolution for the signal is best for large $\Apr$ masses and degraded for a “light” $\Apr$ ($m_\Apr < 50 $ MeV).

\subsection{Expected background}
All processes resulting in a single $\gamma$ hitting the calorimeter represent the background for the experiment, the most relevant being $bremsstrahlung$ and the $e^+ e^-$ annihilation processes in two and three photons. In order to reduce the bremsstrahlung background, the proposed detector features an active veto system composed of plastic scintillating bars: positrons losing energy via bremsstrahlung in the target are detected in the vetoes,  rejecting the event. However the high bremsstrahlung rate is an issue for this class of experiments, limiting the maximum viable beam current. To evaluate this background, a full GEANT4~\cite{Agostinelli:2002hh} simulation of the positron beam impinging on the target, based on the PadmeMC  simulation program \cite{Leonardi:2017lxh},
has been performed. For all bremsstrahlung photons reaching the ECAL, the missing mass has been computed, accounting for the assumed detector angular and momentum resolution. %The total rate of expected bremsstrahlung events for positron on target (POT) was scaled accounting for the effect of the veto system.

The $e^+e^- \rightarrow \gamma \gamma$ and $e^+e^- \rightarrow \gamma \gamma \gamma$ annihilation processes can produce background events whenever only one of the produced photons is detected in the ECAL. This contribution to background has been calculated as follows. Events have been generated directly using CALCHEP, which provided also the total cross sections for the processes. As in the case of bremsstrahlung, the missing mass spectrum was computed for events with a single photon hit in the ECAL. 
\begin{figure}[t!]
    \centering
     \includegraphics[width=.45\textwidth]{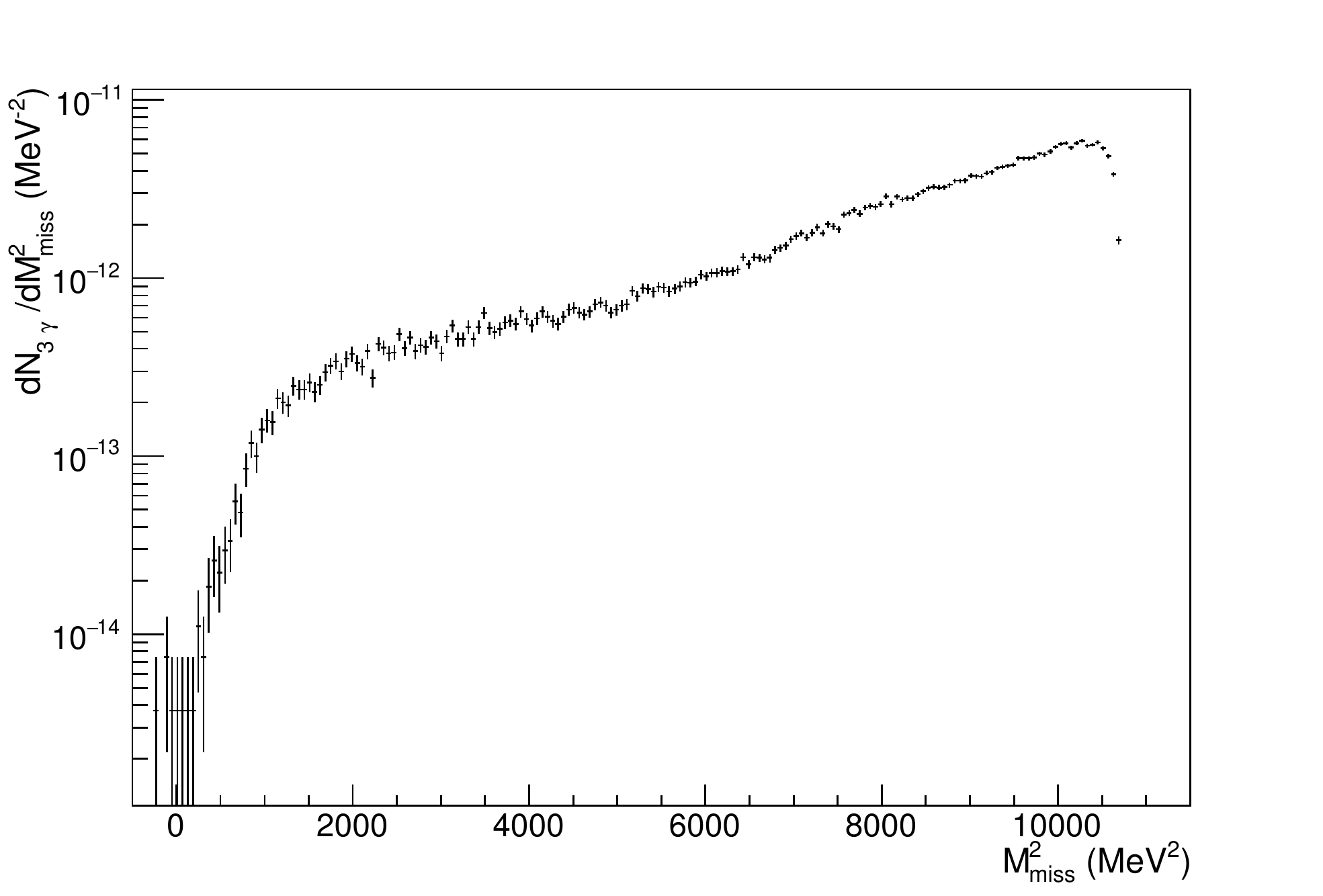}
    \caption{\label{fig:trigamma} Differential missing mass spectrum from positron annihilation into three photons events, normalized to single positron on target. 
    %Computed missing mass spectrum from positron annihilation into three photons events. \textcolor{blue}{Luc: would it be possible to scale it to expected PoT and CS so that it becomes background estimate in a sense? }
    }
\end{figure}
This study proved that, %if a cut of 600 MeV is applied to the measured photon energy, 
if one requires the measured energy to be greater than 600~MeV, the two photon annihilation background  becomes negligible. This is due to momentum conservation: %the closed kinematics of the $e^+e^-$ process: 
asking for only one photon to fall within the ECAL geometrical acceptance translates in a strong constraint on its energy. This argument does not apply to the three-photon annihilation: this process generates an irreducible background for the experiment (see Fig.~\ref{fig:trigamma} for the missing mass spectrum produced by the three-photons annihilation).

\subsection{Experimental Setup}
\begin{figure}[t!]
    \centering
     \includegraphics[width=.5\textwidth]{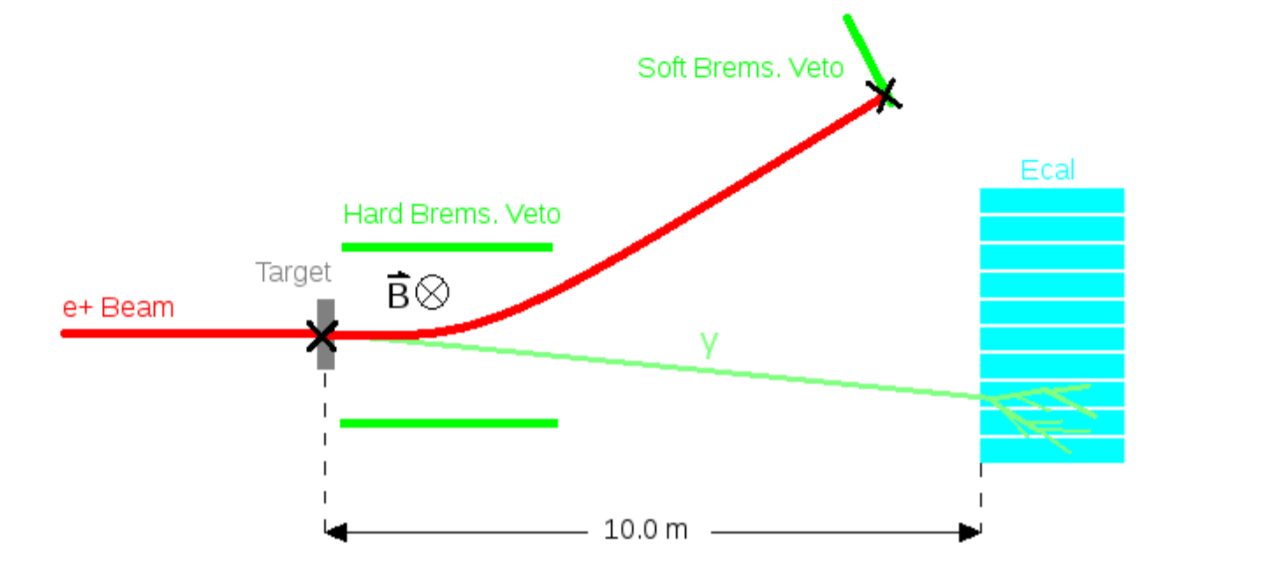}
    \caption{\label{fig:laythin} Layout of the proposed thin target setup.}
\end{figure}

The experimental setup of the proposed measurement is shown in Fig.~\ref{fig:laythin}. The 11 GeV positron beam impinges on a 100-$ \mu m$ thick  target made of carbon, which is a good compromise between density and a low Z/A ratio allowing to reduce the bremsstrahlung rate. A  magnet capable of generating a field of 1 T over a region of 2 m downstream the target bends the charged particles (including  non-interacting positrons) away from the ECAL, placed %10 m
a few meters downstream. The  ECAL is composed of high density  scintillating crystals, arranged in a %50 cm 30 $cm$ radius 
cylindrical shape. High segmentation is necessary to obtain a good angular resolution, critical for a precise missing mass computation, but should however be matched with the Moli\`ere radius of the chosen material.

Crystals of PbWO$_4$, LSO(Ce) and BGO, represent optimal choices,  given the  fast  scintillating time, high-density and short radiation length. Energy resolution, as well as angular resolution, play a crucial role in the missing mass computation; a value of  $\frac{\sigma(E)}{E} = \frac{2\%}{\sqrt{E}}$ has been assumed for this study, consistent with the performance of the 23-cm long PADME BGO detector, corresponding to 20 radiation lengths. Such a depth is indeed needed for achieving this performance, due to longitudinal shower containment. 
Since the small-angle bremsstrahlung high rate would blind the central crystals of the calorimeter, the simplest solution is to imagine a  
% $\sim 20\,\, mm$ radius 
hole at the center of the cylinder. An indicative value of 50~cm for the calorimeter radius and a distance from the target of 10~m were assumed in the evaluation of the projected sensitivity of this experiment. With a spatial resolution of $\sim 5$ mm for the photon impact position, an angular resolution of $0.5$~mrad and a geometrical acceptance of $\sim 50$~mrad are achieved. 
% Assuming a radius of 50~cm and a distance from the target of 10~m, a geometrical acceptance of $\sim 50$~mrad is achieved. 

With a view to adapting the existing PADME detector to perform this measurement -- as thoroughly discussed below -- the assumed acceptance and resolution can be achieved with the 30 cm radius PADME calorimeter, placed at a 6~m distance from the target. In PADME, with a crystal front-face of 20$\times$20 mm$^2$, a spatial resolution of $\sim 3.5$ mm has been measured (significantly better than 20 mm~$/\sqrt{12}$). At 6~m distance this corresponds to an angular resolution of $0.5$~mrad and an acceptance of $\sim 50$~mrad, consistent with the assumed parameters.
%\textcolor{blue}{Paolo: Vedi sopra, invece di 1x1 ho fatto il discorso con 2x2, anche perché 1x1 non matcha il raggio di Moliére di nessuno dei materiali citati, comunque si può spostare il discorso nel paragrafetto su PADME}

Besides the ECAL, the experimental setup includes a veto system  to reduce the bremsstrahlung background. Following the layout of the PADME experiment, the vetoes are composed of plastic scintillator bars. Whenever the primary $e^+$ loses energy via bremsstrahlung in the target, its trajectory is bent by the magnetic field and it impinges on the veto bars, rejecting the event. For the sake of this study, a $99.5\%$ veto efficiency  has been considered. This assumption is proven realistic by the performance of the existing PADME experiment veto system~\cite{Raggi:2014zpa}.
Although this option was not considered in this study, further suppression of the background can be achieved by placing a photon detector, much faster than the main calorimeter, covering its central hole. Such a fast calorimeter would also help in the reduction of $\gamma\gamma$ and $3\gamma$ events with one or two photons lost. In the case of PADME a 5$\times$5 matrix of 3$\times$3 cm$^2$ PbF$_2$ crystals is used. The Cherenkov light from showers is readout by fast photomultipliers, providing a $\sim$ 2~ns double pulse separation (to be compared with $\sim$300~ns decay time of the BGO). 

%\vspace{-1cm}

\subsection{Positron beam requirements}
As already mentioned, the $\Apr$ mass range that the proposed thin target experiment can explore  is strictly constrained by the available energy in the center of mass frame. In this respect, a 11 GeV positron beam would allow extending significantly the $\Apr$ mass range with respect to other similar experiments, up to $\sim106$ MeV$/c^2$.  Being the $e^+e^- \rightarrow \gamma \Apr$ annihilation a rare process, the sensitivity of the proposed search depends on the number of positron on target (POT) collected. In this setup, the maximum current is constrained by the bremsstrahlung rate on the ECAL innermost crystals. Therefore, a continuous beam structure is preferable. In this study, a continuous $100$-$\rm nA$ beam has been considered, resulting in a manageable $\sim200$-$\rm kHz$ rate per crystal in the inner ECAL. In this configuration, $10^{19} $ POT can be collected in $180$ days, covering a new region in the $\Apr$ parameter space. In the event  that the available beam current is lower than $100 \,\,\rm nA$, a similar result can be obtained increasing the target thickness, at the price of a higher background due to multiple scattering.  

The computation of the missing mass requires a precise knowledge of the primary positron momentum; this translates to certain requirements in terms of the quality of the beam. Here, a energy dispersion $\frac{\sigma_{E_{Beam}}}{E_{Beam}} < 1\%$ and an angular dispersion $\theta_{Beam} < 0.1$~mrad of the beam have been considered. With these assumptions, the missing mass resolution is dominated by the ECAL performance, with a negligible contribution from the beam dispersion.

%\textcolor{red}{Mauro: commenti su PADME da mettere in un paragrafo}

\subsection{Reuse of the PADME components}

It's also interesting to investigate the possibility of re\-using 
the existing PADME experimental apparatus as the starting point 
for the new thin target experiment at the CEBAF accelerator. 
In this paper we try to shortly review which part of the 
apparatus could be directly reused, and which will need to be 
adapted to the different beam conditions. 

The PADME target can be easily transferred and installed in the CEBAF 
accelerator, while the option of a ticker target will simplify the design and it 
is easily achievable.
The PADME electromagnetic calorimeter performance is adequate with the requirements for the thin target experiment: in addition to the excellent energy resolution, $<2\% \sqrt{E}$~\cite{Raggi:2016ews}, and spatial resolution, $\sim$ 3.5 mm, single BGO crystals are capable of tolerating rates in excess of 2 MHz. The increased energy of the beam, from 0.5 to 11 GeV, would improve the energy resolution, but will also enhance the contribution of longitudinal shower containment to the resolution with respect to the stochastic term. The overall effect should not degrade the resolution significantly, due to the sufficient total depth of $\sim$20 $X_0$. 

The small angle calorimeter will also profit by the much higher energy of the impinging photons, but will suffer more the longitudinal 
leakage, being only 15 $X_0$ long. This will not compromise its use as
photon veto, while performance as calorimeter, for improving $2\gamma$ and $3\gamma$ acceptance, needs to be evaluated.

The charged veto system will certainly require a different geometrical assembly, both due to the need of a longer magnet and the different boost, but the technology and front-end electronics can be reused.

The trigger and DAQ system of the PADME experiment \cite{Leonardi:2017ocd} was built to operate at a rate of 50 Hz as imposed by the repetition rate of the DA$\Phi$NE LINAC. Currently, PADME is operated in trigger-less mode, i.e. digitizing all channels of the detectors every single beam bunch, typically in a 1 $\mu$s window (1024 samples at 1 Gsample/s).
Of course such a system cannot be used with a continuous beam structure, so that a new trigger and DAQ system need to be designed and built.

%\textcolor{blue}{Paolo: ho aggiunto info su PADME prima e rivisto il paragrafetto}

\section{Positron annihilation on a thick active target}

\subsection{Signal signature and yield}

The cross section for LDM production through positron annihilation on atomic electrons, $e^+ e^- \rightarrow \Apr \rightarrow \chi \overline{\chi}$, is characterised by a resonant shape~\cite{pdg}:
\begin{equation}
    \sigma= \frac{4\pi\alpha_{EM}\alpha_D \varepsilon^2}{\sqrt{s}}
    \frac{q(s-4/3q^2)}{(s-m^2_\Apr)^2+\Gamma_\Apr^2m^2_\Apr}
     \; \; ,
\end{equation}
where $s$ is the $e^+$ $e^-$ system invariant mass squared, $q$ is the $\chi-\overline{\chi}$ momentum in the CM frame, and $\Gamma_\Apr$ is the $\Apr$ width. The kinematics of the $e^+ e^- \rightarrow \chi \overline{\chi}$ reaction in the \textit{on-shell scenario} ($m_\Apr > 2 m_\chi$) is strongly constrained by the underling dynamics. Since the $\Apr$ decays invisibly, its energy is not deposited in the active target, and the corresponding experimental signature is the presence of a peak in the missing energy ($E_{miss}$) distribution, whose position depends solely on the $A^\prime$ mass through the kinematic relation
\begin{equation}
    m_\Apr=\sqrt{2m_e E_{miss}} \; .
\end{equation}

\begin{figure}[t!]
    \centering
     \includegraphics[width=.41\textwidth
     ]{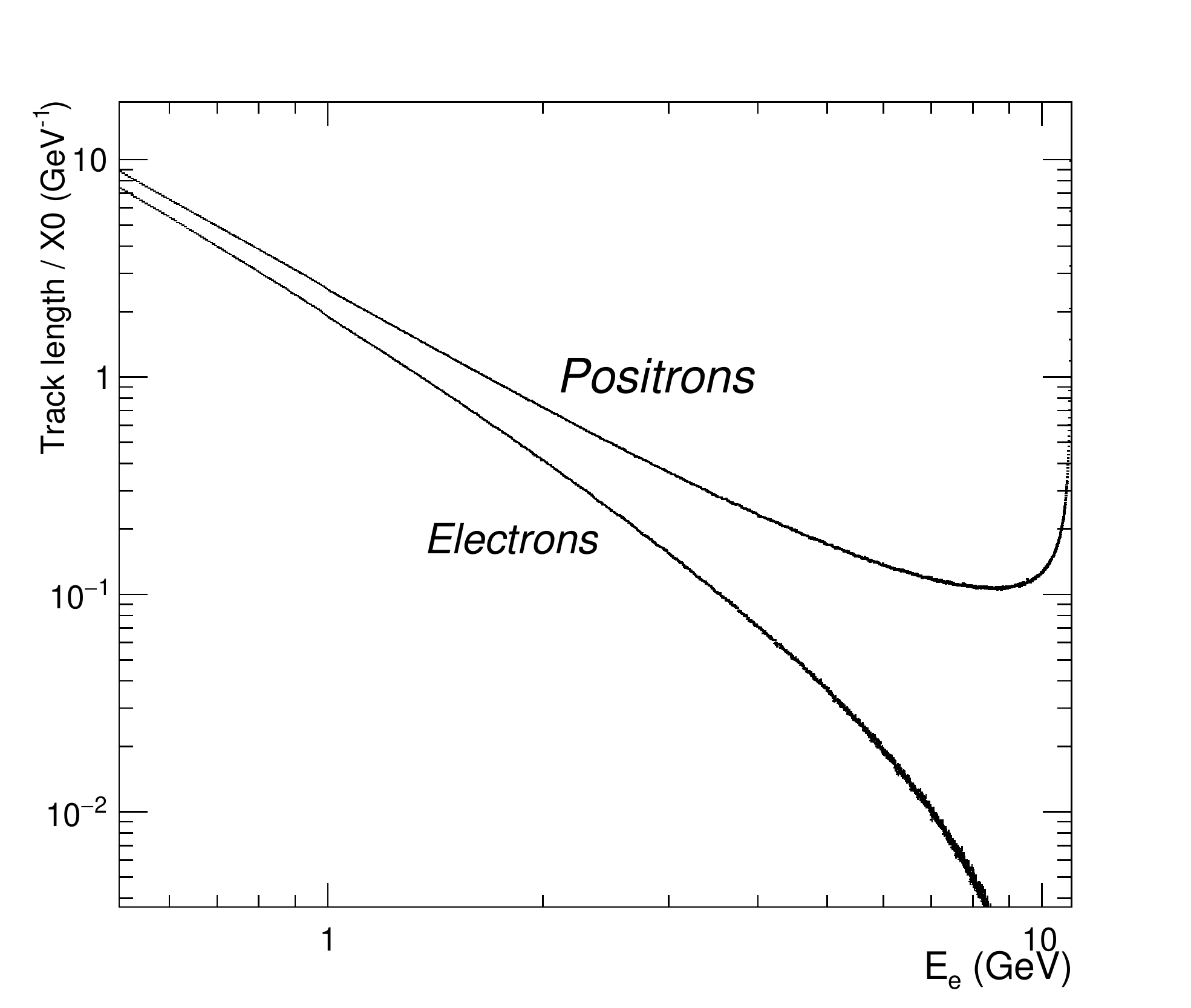}
    \caption{\label{fig:trackLength} Differential positrons track length distribution, normalized to the radiation length, for a 11 GeV $e^+$ beam impinging on a thick target. For comparison, the same distribution in case of an impinging electron beam is reported.}
\end{figure}

For a given $\Apr$ mass, the expected signal yield is:
\begin{equation}\label{eq-prod}
N_{s}= n_{POT}\frac{N_A}{A} Z \rho \int_{E_{miss}^{CUT}}^{E_0} dE_e \,\, T_+(E_e)\,\sigma(E_e) \; ,
\end{equation}
where $A$, $Z$, $\rho$, are, respectively, the target material atomic mass, atomic number and mass density, $E_0$ is the primary beam energy, $N_A$ is Avogadro's number, $\sigma(E_e)$ is the energy-dependent production cross section, $n_{POT}$ is the number of impinging positrons and ${E_{miss}^{CUT}}$ is the missing energy cut. Finally, $T_+(E_e)$ is the positrons differential track-length distribution~\cite{Chilton}, reported in Fig.~\ref{fig:trackLength} for a 11 GeV positron beam.

\subsection{Positron beam requirements}

A missing energy measurement requires that the intensity of the primary positron beam is low enough so that individual $e^+$ impinging on the active target can be distinguished. At the same time, the beam current has to be large enough  to accumulate a sizeable number of positrons on target (POT). For example, a positron beam with a time structure corresponding to 1 $e^+/\mu$s can accumulate more than $10^{13}$ POT/year, with an average time interval between positrons of 1 $\mu$s.

\begin{figure}[t!]
    \centering
     \includegraphics[width=.43\textwidth]{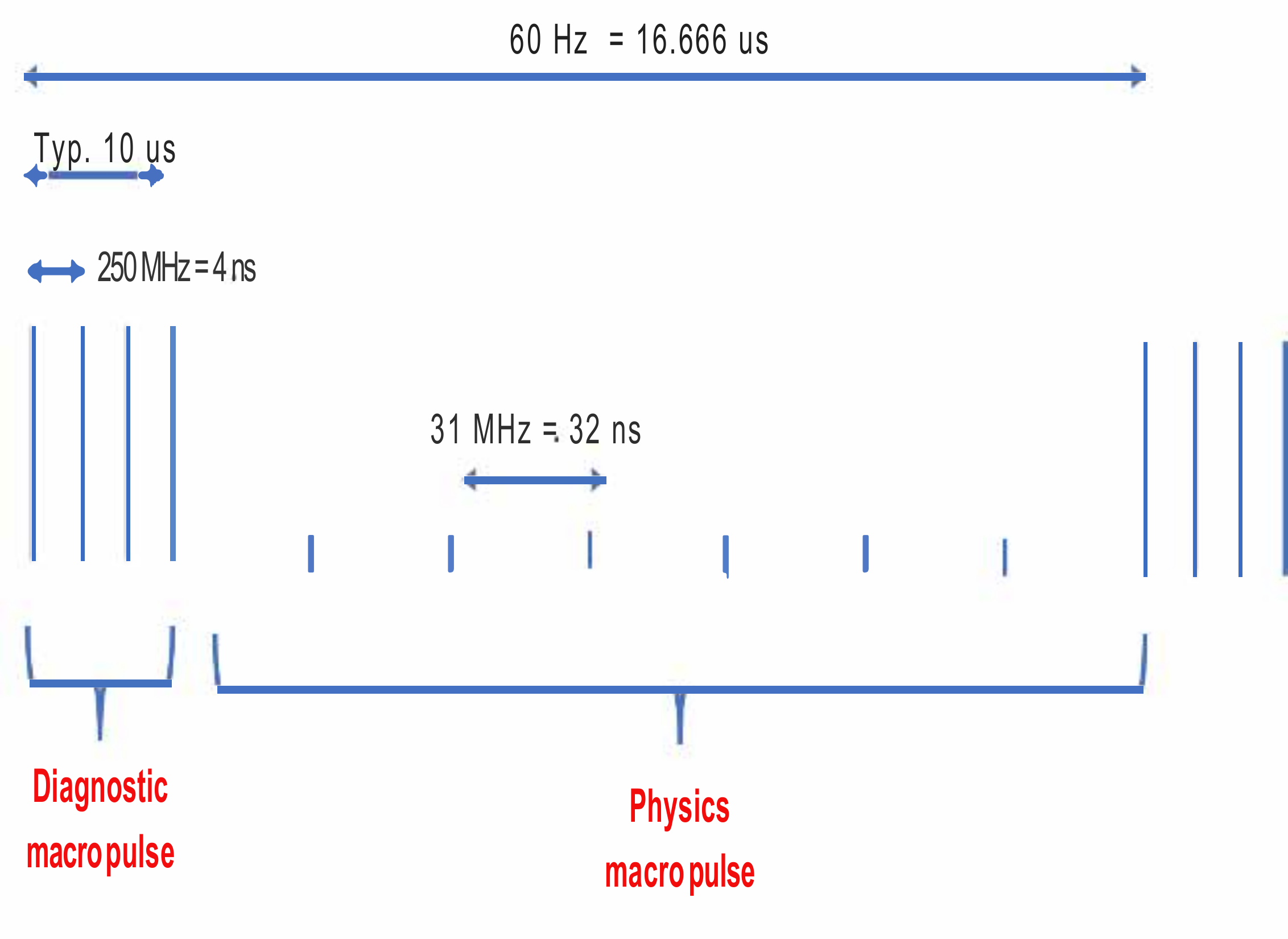}
    \caption{\label{fig:timestructure} Simplified scheme of the $e^+$ beam time structure for the thick-target measurements, see text for details.}
\end{figure}

This specific time structure is challenging for the proposed CEBAF $e^+$ operations. In particular, the low beam current, $\sim 0.1$ pA, is incompatible with the standard beam diagnostic tools that are employed to properly steer and control the  CEBAF beam. Therefore, the following ``mixed operation mode'' is currently being considered for the experiment (see also Fig.~\ref{fig:timestructure})~\cite{GramesVoitier}. A 10--$\mu$s long 100 nA \textit{diagnostic macro-pulse} is injected in the CEBAF accelerator with a 60 Hz frequency. This results to an average current of 60 pA, with a peak current large enough to enable proper operation of the beam diagnostic systems. In between every pulse, low intensity \textit{physics pulses}, populated \textit{on average} by less than 1 $e^+$, are injected at higher frequency. 

This challenging operation scheme can be realized  using an ad-hoc laser system at the injector. With dedicated R$\&$D, it would be possible to design and construct a system capable of injecting fast bunches at 31.25 MHz - i.e. one bunch every 32 ns. Since the (discrete) number of positrons per bunch follows a Poissonian statistical distribution, the time interval between $e^+$ can be further increased by reducing the average bunch population, by adjusting the laser intensity. A $\sim $500 ns spacing between positrons can be obtained by using an average laser power of 0.05 $e^+$/bunch.
The experiment will acquire data only during low-intensity pulses, ignoring the 10 $\mu$s long high current periods. However, if all these positrons would impact on the detector, the average rate of $\sim3.7$ $10^8$ $e^+$/s would result in a very large radiation dose deposited in the active target. To avoid this, we plan to install in front of the detector a fast magnetic deflector, synchronized to the beam 60 Hz frequency, in order to transport the positrons belonging to the high-current pulses to a suitable beam-dump, avoiding their impact on the detector.

In summary, the proposed CEBAF operation mode would allow to obtain a beam with positrons impinging on the detector on average every $\sim 500$ ns, compatible with the accelerator control and diagnostic system.  It should be remarked that this technical solution requires R$\&$D activities, that are already (partially) planned in the contest of EIC accelerator development. In the following, we will present the sensitivity to DM considering $10^{13}$ POT accumulated in one year of run.

\subsection{Experimental setup}
\begin{figure}[t!]
    \centering
     \includegraphics[width=.41\textwidth]{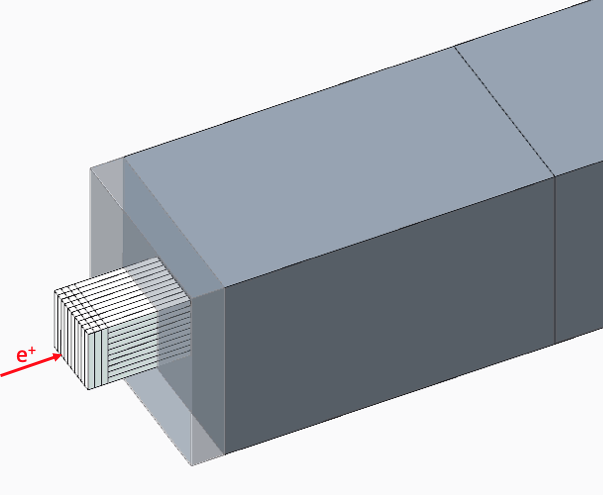}
    \caption{\label{fig:thick-target} Schematic layout of the active thick-target experimental setup, with the ECAL (white) followed and surrounded by the HCAL (grey). The semi-transparent portion of the HCAL in front is that installed all around the ECAL.}
\end{figure}

The layout of the proposed measurement is schematically reported in Fig.~\ref{fig:thick-target}. It includes a homogeneous electromagnetic calorimeter (ECAL) acting as  a thick target to measure the energy of each impinging positron, and a hadron detection system (HCAL) installed around and downstream the active target to measure long-lived (neutrons/$K_L$) or highly penetration (muons/charged pions) particles escaping from the ECAL.

The \textit{preliminary} ECAL design foresees a 28 radiation lengths detector, made as a $10\times10$ matrix of PbWO$_4$ crystals of dimensions of $20\times20\times250$ mm$^3$. Three layers of crystals are added in front, with the long axis oriented perpendicular to the beam direction, to act as a preshower, resulting in a total calorimeter length of 35$X_0$. The choice of PbWO$_4$ material is motivated by its fast scintillating time ($\tau \simeq 30$ ns), well matched to the expected hit rate, its high-density, resulting in a compact detector, and its high radiation hardness. The total calorimeter length was selected to limit below ${\sim}10^{-13}$ per POT the probability that any particle from the developing cascade, in particularly photons, escape the detector faking a signal. The transverse size, was chosen to provide measurements of the shower transverse profile and to optimize the optical matching with the light sensor. The total front face size (20x20 cm$^2$) is large enough to avoid transverse energy leakage affecting the detector resolution. SiPMs will be used to collect scintillation light from the crystals. The use of these sensors has never been adopted so far in high-energy electromagnetic calorimetry with PbWO$_4$ crystals, and requires a careful selection of the corresponding parameters. First measurements on PbWO$_4$ crystals with 6x6 mm$^2$ devices having a 25 $\mu$m pixel size show a light yield of $\sim$ 1 phe / MeV, compatible with the experiment requirements (energy resolution and dynamic range). The expected radiation dose for the detector, for positrons impinging on the calorimeter every 500 ns and assuming an overall beam availability of 50$\%$ is, at maximum, $\sim$ 350 rad/h, corresponding to the central crystals. This large value, comparable to the maximum dose in the CMS PbWO$_4$ electromagnetic calorimeter~\cite{Adzic:2009aa,Chatrchyan:2008aa}, calls for a careful calorimeter design and for the identification of procedures to mitigate any possible radiation damage during detector operation. These include varying the beam impact point on the detector to distribute  the radiation dose across crystals, as well as annealing crystals during off-beam operations, exploiting both thermal annealing and light-induced processes~\cite{DORMENEV20101082,Fegan:2015via}. 

\begin{figure*}[t!]
    \centering
     \includegraphics[width=.45\textwidth]{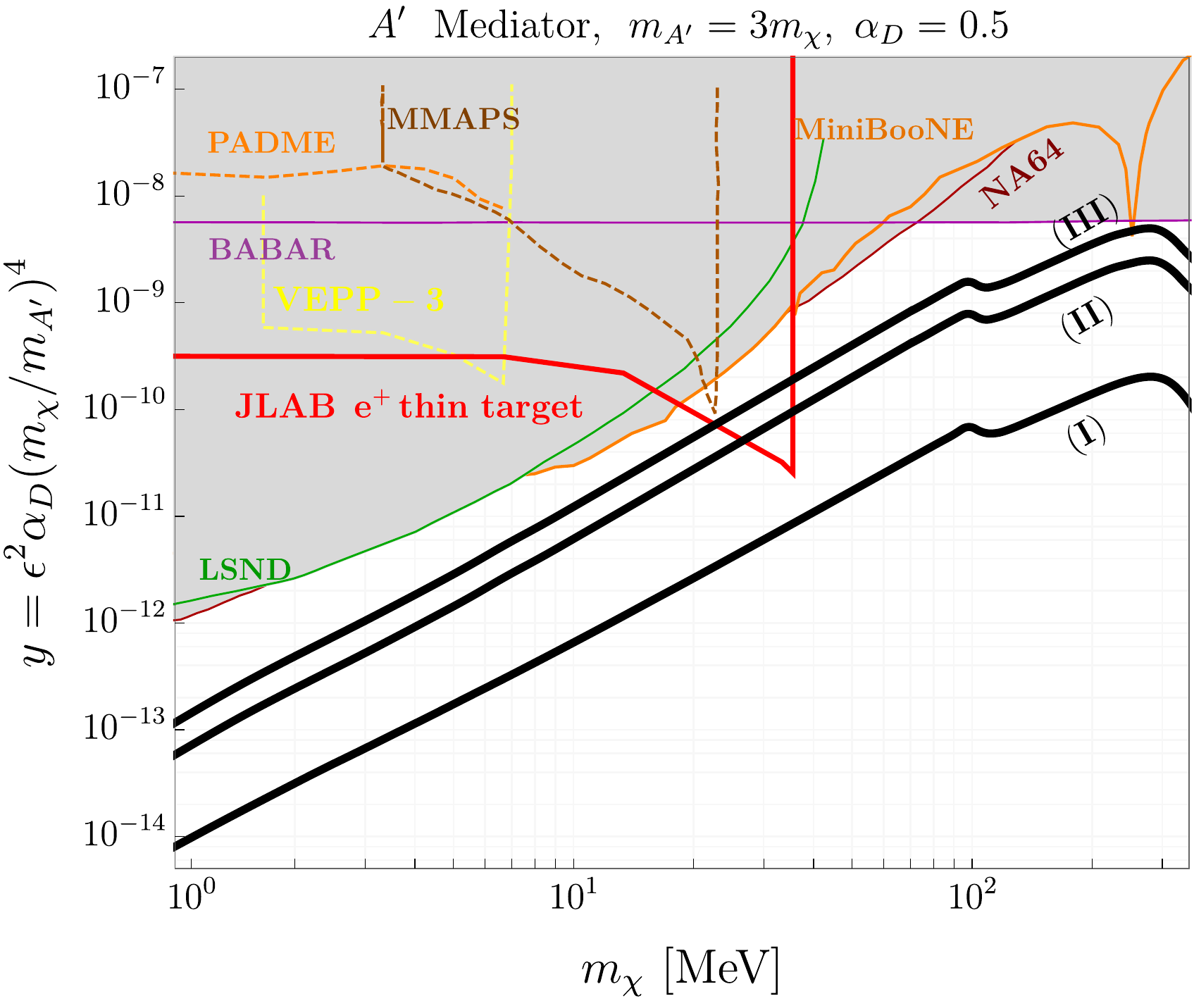}
     \quad
     \includegraphics[width=.45\textwidth]{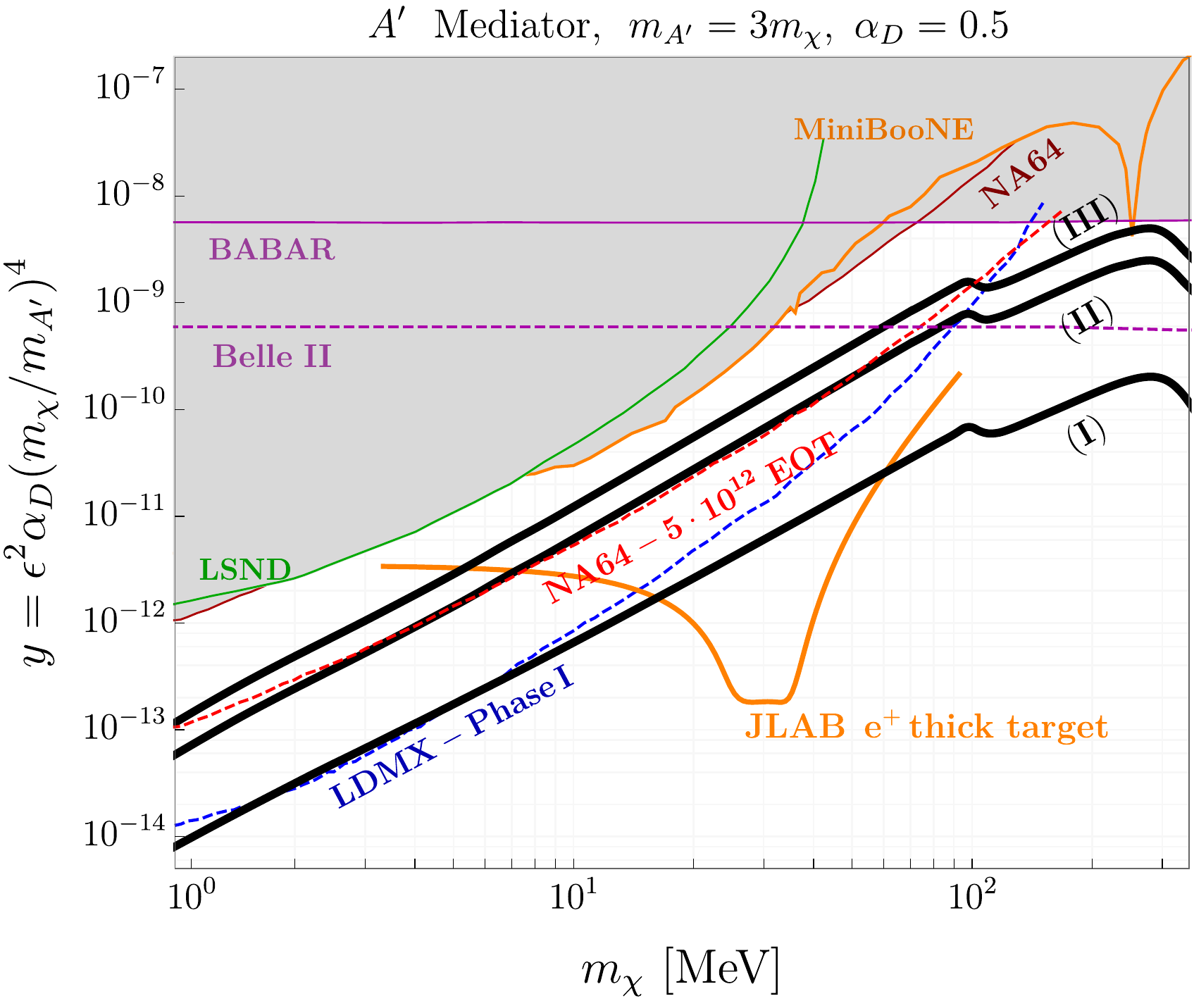}
    \caption{\label{fig:sensitivity} The expected sensitivity for the thin-target (red, left) and thick-target (orange, right) measurements, compared to existing exclusion limits (grey area~\cite{Lees:2017lec,NA64:2019imj,Aguilar-Arevalo:2018wea}) and projections for future efforts (dotted lines: PADME~\cite{Raggi:2015gza}, MMAPS~\cite{Alexander:2017rfd}, VEPP-3~\cite{Wojtsekhowski:2017ijn}, Belle-II~\cite{Kou:2018nap}, LDMX~\cite{Berlin:2018bsc} and NA64~\cite{Beacham:2019nyx}). The black lines are the thermal targets for pseudo-Dirac fermion LDM (I), Majorana fermion LDM (II) and  
    elastic and inelastic scalar LDM (III).}
\end{figure*}

The main requirement for the HCAL is the hermeticity to long-lived particles exiting from the ECAL. From a Montecarlo simulation of this setup, the probability of having one or more high-energy (${\gtrsim} 1$ GeV) hadron leaving the active target is ${\sim} 10^{-4}$ per POT. This calls for a HCAL inefficiency of $10^{-10}$ or lower. The \textit{preliminary} detector design ueses a modular iron/scintil\-lator inhomogeneous calorimeter, with a length corresponding to approximately 25 nuclear interaction lengths, partially surrounding the active target to avoid any particle leakage from the calorimeter lateral faces. 

\subsection{Measurement and analysis strategy}

The experiment will be characterised by a very high measurement rate, dominated by events with full energy deposition in the calorimeter.
To cope with this, the data acquisition system will be configured to record only events with a significant ($\gtrsim 1$ GeV) energy loss in the calorimeter. From a preliminary estimate, the expected trigger rate will be ${\sim}20$ kHz, for a primary beam impinging with 2 MHz frequency on the detector. This minimum-bias condition will be initially studied with Montecarlo simulations, to evaluate the efficiency and confirm that no distortions to the experiment physics outcome are introduced. In parallel to the main production trigger, prescaled trigger conditions will be implemented to save %a large-enough sample of
full-energy events for calibration and monitoring.
A blind approach to data analysis will be followed. First, events in the signal region, based on a preliminary choice of the calorimeter and hadron detection system energy cuts, will be excluded from the analysis. Then, the expected number of backgrounds will be evaluated using both Montecarlo simulations and events in the neighborhood of the signal region, in order to identify an optimal set of selection cuts for the signal that maximize the experiment sensitivity~\cite{Cowan:2010js}. Finally, the signal region will be scrutinized. 

 \vspace{-.2cm}

\section{Results}

The 90$\%$ CL sensitivity of the two proposed measurements is shown in Fig.~\ref{fig:sensitivity}, compared with current exclusion limits (grey areas) and expected performance of other missing-energy / missing-mass future experiments (dashed curves). On the same plot, we show the thermal targets for significant variations of the minimal LDM model presented in the introduction: elastic and inelastic scalar LDM (I), Majorana fermion LDM (II) and pseudo-Dirac fermion LDM (III). For the thin-target effort, the red curve reports the sensitivity estimate based on the realistic backgrounds that have been discussed before. For the thick-target case,  the orange curve refers to the ideal case of a zero-background measurement, considering a 5.5 GeV missing energy threshold, and assuming an overall 50$\%$ signal efficiency. 
This hypothesis, following what was done in similar experiments~\cite{LDMX,NA64:2019imj}, will be investigated with Montecarlo simulations during the future experiment design phase.

%\begin{figure}
%    \centering
%     \includegraphics[width=.48\textwidth]{reach.pdf}
%    \caption{\label{fig:sensitivity} The expected sensitivity for the thin-target (red) and thick-target (orange) measurements, compared to existing exclusion limits (grey area) and projections for future efforts (dotted lines). The black lines are the thermal targets for  elastic and inelastic scalar LDM (I), Majorana fermion LDM (II), and pseudo-Dirac fermion LDM (III).}
%\end{figure}

\subsection{Complementarity of the two approaches}

The two measurements that we presented in this document are characterized by different sensitivities and design complexity. They can be considered as two complementary experiments facing the light dark matter physical problem, and as such we foresee a comprehensive LDM physical program at JLab with both of them running, but with different time schedules.

With the availability of a 100-nA, 11-GeV positron beam at JLab, the thin-target experiment can start almost immediately, since no demanding requirements on the beam are present. The conceptual design is already mature, being based on realistic Monte Carlo simulations. Furthermore, the detector can be based on an already-existing and working setup, the PADME experiment at LNF~\cite{Raggi:2014zpa}. As discussed before, the possibility of installing PADME at JLab, benefiting from both the existing equipment and our experience is a compelling possibility, allowing us to run successfully the thin-target measurements from day one. 

Meanwhile, we propose starting the necessary R$\&$D activity in preparation to the thick-target measurement, exploiting synergic activities at the laboratory in the context of the EIC program. 
The goal is to be ready to start the measurements on a time scale of few years after the beginning 
of the $e^+$ program at JLab.

 \vspace{-.2cm}

\section{Conclusions and outlook}

In this document, we have discussed two complementary strategies to explore the phenomenology of 
dark sectors by exploiting a future $e^+$ beam at JLab. The unique properties of this facility -- the high energy, the high intensity and the versatile operation mode -- will allow these two experimental approaches to investigate vast unexplored regions in the parameters space, beyond those  
covered by current or planned experiments.
An  experimental program more comprehensive than the one discussed here 
can also be  envisaged, 
% a comprehensive experimental program, 
which would include  dedicated measurements to investigate more in depth some specific LDM scenarios. % including the most important variations of the vanilla model here discussed. 
Possible efforts include, for example, 
% a beam-dump experiment with a positron beam to investigate both the visible and invisible LDM % scenario~\cite{PhysRevLett.121.041802,PhysRevD.98.015031}, as well as a 
dedicated measurements aimed to scrutinize the explanation in terms of a dark boson of 
the recently reported $^8$Be and $^4$He anomalies~\cite{PhysRevLett.116.042501,Krasznahorkay:2019lyl}.

In summary, \textbf{the availability of a positron beam will make JLab the premier facility for exploring the dark sector, and the proposed experimental program will allow for the confirmation or rejection of the LDM hypothesis by covering the thermal targets in a wide region in the parameters space.} \newline

This material is based upon work supported by the U.S. Department of
Energy, Office of Science, Office of Nuclear Physics under contract
DE-AC05-060R23177.

%DE$-$⁠AC05$-$⁠06OR23177.

% Although not discussed in this document, w

\bibliographystyle{unsrt}  
%\bibliographystyle{spphys}  
% \bibliographystyle{spmpsci}      % mathematics and physical sciences
% \bibliographystyle{abbrv}       % APS-like style for physics

%\bibliography{JPOS_EPJ}

\end{document}